\documentclass[11pt]{article}
\usepackage{path,graphicx}
\usepackage{a4,amssymb, dsfont}
\usepackage{authblk}
\usepackage{amsmath}
\usepackage{cite}
\usepackage[]{algorithm2e}
\usepackage[margin=1.2in]{geometry}
\usepackage[T1]{fontenc}
\usepackage[utf8]{inputenc}
\usepackage{authblk}
\usepackage[justification=centering]{caption}
\newtheorem{definition}{Definition}[section]
\newtheorem{lemma}[definition]{Lemma}
\newtheorem{theorem}[definition]{Theorem}

\newtheorem{corollary}[definition]{Corollary}

\newtheorem{remark}[definition]{Remark}

\newcommand{\argmax}{\mathop{\mathrm{argmax}}}
\newcommand{\argmin}{\mathop{\mathrm{argmin}}}
\newcommand{\uar}{\leftarrow_{u.a.r.}}
\title{(Biased) Majority Rule Cellular Automata\footnote{This paper is an extended version of work published in \cite{gartner2017color}.}}
\author{Bernd G\"artner\thanks{gaertner@inf.ethz.ch}}
\author{Ahad N. Zehmakan\thanks{abdolahad.noori@inf.ethz.ch }}
\affil{Department of Computer Science, ETH Zurich}

\providecommand{\keywords}[1]{\textbf{\textit{Key Words:}} #1}
\date{} 
\begin{document}
\maketitle

\begin{abstract}
Consider a graph $G=(V,E)$ and a random initial vertex-coloring, where each vertex is blue independently with probability $p_{b}$, and red with probability $p_r=1-p_b$. In each step, all vertices change their current color synchronously to the most frequent color in their neighborhood and in case of a tie, a vertex conserves its current color; this model is called \textit{majority model}. If in case of a tie a vertex always chooses blue color, it is called \textit{biased majority model}. We are interested in the behavior of these deterministic processes, especially in a two-dimensional torus (i.e., cellular automaton with (biased) majority rule). In the present paper, as a main result we prove both majority and biased majority cellular automata exhibit a threshold behavior with two phase transitions. More precisely, it is shown that for a two-dimensional torus $T_{n,n}$, there are two thresholds $0\leq p_1, p_2\leq 1$ such that $p_b \ll p_1$, $p_1 \ll p_b \ll p_2$, and $p_2 \ll p_b$ result in monochromatic configuration by red, stable coexistence of both colors, and monochromatic configuration by blue, respectively in $\mathcal{O}(n^2)$ number of steps.   
\end{abstract}
\section{Introduction}
\label{introduction}
Suppose that in a community, people have different opinions on a topic of common interest. Through social interactions, individuals learn about the opinions of others, and as a result may change their own opinion. The goal is to understand and possibly predict how opinions spread in the community. There are numerous mathematical models for such a situation; a very simple deterministic one is the following: the community is modeled as a graph, with edges corresponding to possible interactions between individuals. Opinions spread in rounds, where in each round, each individual adopts the most frequent opinion in its neighborhood. If in case of a tie an individual stays with him/her opinion, it is named \emph{majority model}, but if in case of a tie s/he always adopts a specific opinion, the process is called \emph{biased majority model}. These two natural updating rules have various applications, for example in data redundancy \cite{peleg2002local}, distributed computing \cite{perron2009using}, modeling biological interactions \cite{cardelli2012cell}, resource allocation for ensuring mutual exclusion \cite{peleg2002local}, distributed fault-local mending \cite{peleg2002local}, and modeling diffusion of two competing technologies over a social network \cite{fazli2014non}.

Scientists from different fields, from Spitzer \cite{spitzer1991interaction} to a recent paper by Mitsche \cite{mitsche2015strong}, have attempted to study the behavior of these natural updating rules, especially on the two-dimensional torus where the (biased) majority rule can be interpreted as a cellular automaton. Some theoretical and experimental results concerning its behavior have been obtained, which will be discuss in detail in Section \ref{Prior Works}; of particular interest is the \emph{consensus time}, the time after which the process reaches a periodic sequence of states (which must eventually happen, as the process is deterministic and has finite state space). Also, one would like to understand how these ``final'' states looks like, depending on the initial distribution of opinions. For example, what are conditions under which some opinion is eventually taken up by all the individuals? 

\begin{figure}[h]
\begin{center}
\includegraphics[width=0.9\textwidth]{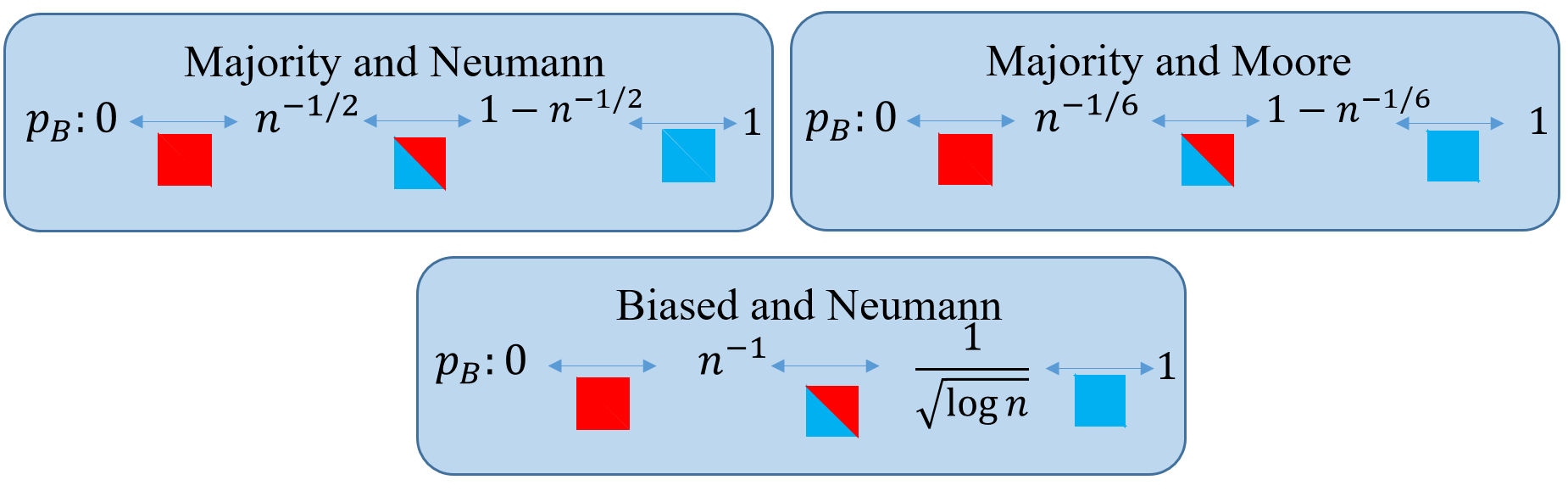}
\caption{The threshold values for both majority and biased majority cellular automata in Moore and Neumann neighborhoods (by $p_b \ll 1-f(n)$ (similarly $p_b \gg 1-f(n)$) we mean $p_r \gg f(n)$ ($p_r\ll f(n)$) for $f(n)=n^{-1/2}$ and $n^{-1/6}$) \label{fig 2}}
\end{center}
\end{figure}

In this paper, we first present some results in general graphs $G=(V,E)$, regarding consensus time, eternal sets (sets of nodes that guarantee the survival of an opinion that they have in common), and robust sets (sets of nodes that will never change an opinion that they have in common). Building on them, our main contribution is for the case where $G$ is an $n\times n$ torus, with $4$-neighborhoods (Neumann neighborhood), or with $8$-neighborhoods (Moore neighborhood). As mentioned, we study the case of two opinions (modeled by vertex colors blue and red), with an initial coloring that assigns blue to every vertex independently with a probability $p_b=p_b(n)$, and red otherwise ($p_r=1-p_b$). It is proven that both majority cellular automata and biased majority cellular automata exhibit a threshold behavior with two phase transitions in Neumann neighborhood. In the torus $T_{n,n}$ with Neumann neighborhood with high probability, majority model results in a monochromatic generation with red, stable coexistence of both colors, and monochromatic generation with blue for $p_b\ll n^{-1/2}$, $n^{-1/2} \ll p_b, p_r$, and $p_r\ll n^{-1/2}$, respectively in $\mathcal{O}(n^2)$ number of steps. Furthermore, it is proved that for Neumann neighborhood and the biased majority model, $p_b \ll n^{-1}$, $n^{-1} \ll p_b \ll 1/\sqrt{\log n}$, and $1/\sqrt{\log n} \ll p_b$ result in final red monochromatic configuration, stable coexistence of both colors, and final complete occupancy by blue, respectively in $\mathcal{O}(n^2)$ steps with high probability. We also prove that majority cellular automata show a threshold behavior in the case of Moore neighborhood. Figure \ref{fig 2} summarizes the mentioned threshold behaviors.\footnote{$f(n) \ll g(n)$ means $f(n)\in o(g(n))$ for two functions $f(n)$ and $g(n)$.}

These results not only are important in their own right, but also answer two important questions. Firstly, prior empirical research \cite{molofsky1999local, oliveira2009some, de1992isotropic, shao2009dynamic} demonstrate that majority cellular automata should show a threshold behavior i.e. if the concentrations of vertices holding the same color is above a certain threshold, the color will survive in all upcoming configurations, otherwise the color might disappear in a few steps. Our results concerning majority cellular automata prove the aforementioned empirical observation and determine the threshold values and the consensus time of the process precisely. Secondly, Schonmann \cite{schonmann1990finite} proved that in biased majority cellular automata with Neumann neighborhood $p_b \gg 1/\sqrt{\log n}$ outputs final complete occupancy by blue with high probability. We show that this bound is tight; actually, we prove what exactly happens in $p_b \ll 1/\sqrt{\log n}$. Furthermore, we will present the tight bound of $\mathcal{O}(|V|)$ ($|V|$ is the number of vertices) on the consensus time of both automata which has been a point of interest in the prior works.

Intuitively, if the concentration of blue color increases, the chance of its survival will increase and the chance of survival for red color decreases. In other words, by increasing the initial density of a color from 0 to 1, it will go through different phases: (i) very low concentration results in the disappearance of the color (ii) sufficient initial density for both colors outputs the survival of both of them (iii) very high initial concentration of a color results in the final occupancy by the color. Therefore, intuitively and as also prior experimental results have shown, it is not so surprising if one proves that majority and biased majority automata show a threshold behavior with two phase transitions, but the most surprising part is the substantial change in the value of thresholds by switching form the majority model to the biased majority model. In majority cellular automata with Neumann neighborhood, $p_b$ should be very close to 1 to have a high chance of final complete occupancy by blue, but by just changing the tie breaking rule in favor of blue, the process ends up in a blue monochromatic configuration with high probability even for initial concentration very close to 0. Hence, it not only shows the significant impact of the tie breaking rule, but also demonstrates how small alternations in local behavior can result in considerable changes in global behaviors.

To prove that majority cellular automaton has a threshold behavior, we show there exists a blue robust set (a robust set whose all vertices are blue) in starting configuration with high probability for $p_b \gg n^{-1/2}$ ($p_b \gg n^{-1/6}$ in Moore neighborhood) which result in the survival of blue color, but proving $p_b \ll n^{-1/2}$ (similarly $p_b \ll n^{-1/6}$ in Moore) results in a red monochromatic configuration is more difficult. We show in this case with high probability in the initial generation, we can classify the blue cells such that all blue cells in a group behave independently of all other blue cells and the number of blue cells in none of these groups is sufficient for survival. As we will discuss in more detail, this proof technique requires more technical arguments in the case of Moore neighborhood because of switching from 4-neighborhood model to 8-neighborhood model. 

In the case of biased majority cellular automata as we mentioned, Schonmann \cite{schonmann1990finite} proved that $p_b \gg 1/\sqrt{\log n}$ results in final complete occupancy by blue almost surely. To prove, for $n^{-1} \ll p_b \ll 1/\sqrt{\log n}$ in Neumann neighborhood both colors survive, we show there exists at least a blue eternal set (an eternal set whose all vertices are blue) in the initial configuration with high probability, but for proving that red color will never vanish, we need a more complicated argument. For the case of $p_b \ll n^{-1}$, we intuitively have to overcome the same difficulty that we had in the case of majority model; we have to accurately analyze the behavior of the model in several upcoming configurations to show that blue color in this setting will disappear finally.

We also discuss that majority and biased majority cellular automata reach a configuration of period one or two in $\mathcal{O}(|V|)$ steps and these bounds are tight i.e. there are some cases which the process needs $\Theta (|V|)$ steps to stabilize in a configuration of period one or two. We highly depend on the strong results by Poljak an Turzik \cite{poljak1986pre} concerning our results on consensus time and periodicity. 

The layout of the paper is as follows. In the rest of this section, first we introduce majority and biased majority models formally. Then, in Section \ref{Prior Works} we briefly discuss relevant prior research works. In Section \ref{One-dimensional}, the majority model is discussed on cycles (one-dimensional cellular automata) as a simple example before going through two-dimensional cellular automata. In Section \ref{Not-Monochromatic Periodicity in Color War}, we discuss the consensus time and periodicity of the two models on an arbitrary graph $G=(V,E)$. Then, in Sections \ref{majority} and \ref{biased majority}, we focus on the threshold behavior of majority cellular automata and biased majority cellular automata, respectively. 
          
\subsection{Notation, Preliminaries, and (Biased) Majority Model}

Let $G=(V,E)$ be a graph that we keep fixed throughout. For a vertex $v\in V$, $N(v) := \{u\in V: \{v,u\}\in E\}$ is the \emph{neighborhood} of $v$. We also define $\hat{N}(v) := N(v) \cup \{v\}$. Furthermore, for a set $S\subseteq V$, $N_S(v):=N(v) \cap S$ and $N(S):=\bigcup_{v\in S}N(v)$ (similarly $\hat{N}_S(v):=\hat{N}(v) \cap S$ and $\hat{N}(S):=\bigcup_{v\in S}\hat{N}(v)$). 
\\
A \emph{generation} is a function $g:V\rightarrow\{b,r\}$ ($b$ and $r$ represent blue and red, respectively). If $g$ is a constant function, $g$ is called a \emph{monochromatic} generation otherwise it is called a \emph{bichromatic} generation.
$S\subseteq V$ is a \emph{$c$-community} for color $c \in \{b,r\}$ and in  generation $g$ if $\forall v\in S$ $g(v)=c$. For a generation $g$, vertex $v\in V$ and color $c\in\{b,r\}$,
\[N^g_c(v) := \{u\in N(v): g(u) = c\}\] is the set of neighbors of $v$ of color $c$ in generation $g$. We also define \[\hat{N}^g_c(v) := \{u\in \hat{N}(v): g(u) = c\}.\] 
For a set $S\subseteq V$, we define $N_{c}^{g}(S):=\{u\in N(S):g(u)=c\}$ and similarly $\hat{N}_{c}^{g}(S):=\{u\in \hat{N}(S):g(u)=c\}$.
\\
In addition to $g(v)=c$ for a vertex $v\in V$ and $c\in \{b,r\}$, sometimes we also write $g|_S=c$ for a set $S\subseteq V$ which means $\forall v\in S$, $g(v)=c$. 

Given an initial generation $g_0$ such that $\forall v\in V$, $Pr[g_0(v)=b]=p_b$ and $Pr[g_0(v)=r]=p_r$ independently of all other vertices, and $p_b+p_r=1$. Let $\forall i\geq 1$ and $v\in V$, $g_i(v)$ equal to the color that occurs most frequently in $v$'s neighborhood in $g_{i-1}$, and in the case of a tie, $v$ conserves its current color. More formally: 
\[
g_i(v)= \left\{\begin{array}{lll}g_{i-1}(v), &\mbox{if $|N^{g_{i-1}}_b(v)|=|N^{g_{i-1}}_r(v)|$}, \\
\argmax_{c\in\{b,r\}} |N^{g_{i-1}}_c(v)|,&\mbox{otherwise}
\end{array}\right.
\]
The above model is called \emph{Majority Model}. In the same setting by just changing the tie breaking rule, we have \emph{Biased Majority Model}. In biased majority model in case of a tie, a vertex always adopts blue color. More formally:
\[
g_i(v)= \left\{\begin{array}{lll}b, &\mbox{if $|N^{g_{i-1}}_b(v)|=|N^{g_{i-1}}_r(v)|$}, \\
\argmax_{c\in\{b,r\}} |N^{g_{i-1}}_c(v)|,&\mbox{otherwise}
\end{array}\right.
\]
 In the present paper, we discuss the behavior of these two deterministic processes with a random initial coloring.

 Since both models are deterministic and the number of possible generations for a graph $G=(V,E)$ is finite ($2^{|V|}$), they always reach a cycle of generations after finite number of steps. For a graph $G=(V,E)$ and (biased) majority model, the number of steps which the process needs to stabilize (reach a cycle of generations) is called \emph{Consensus Time}. We say a process gets \emph{b-monochromatic} (similarly \emph{r-monochromatic}) if finally it reaches a blue (red) monochromatic generation (all vertices blue (red)), otherwise we say it gets \emph{bichromatic}. Furthermore, for a graph $G=(V,E)$ and (biased) majority model, $Pr[$b-monochromatic$]$ ($Pr[$r-monochromatic$]$) denotes the probability that the graph gets blue (red) monochromatic, and $Pr[$bichromatic$]$=$1-Pr[$b-monochromatic$]$-$Pr[$r-monochromatic$]$ for a random initial coloring in (biased) majority model. 
\begin{remark}
For a graph $G=(V,E)$, we say an event happens with high probability (or almost surely) if its probability is at least $1-o(1)$ as a function of $|V|$.\footnote{Actually, as we will discuss it is mostly $1-e^{-|V|}$.}
\end{remark}

\subsection{Prior Works} 
\label{Prior Works}
 
As mentioned, majority-voting rule has been studied in different literatures because of its importance and applications. Therefore, based on different motivations and from  a wide spectrum of approaches, various definitions of majority rule have been presented, but in general we may classify them into the three following categories.

The first class is the \textit{$\alpha$-monotone model} in which, at each step a vertex becomes blue if at least $\alpha$ of its neighbors are blue, and once blue no cell ever becomes red (in the literature this is also known as \textit{bootstrap percolation}). For example in \cite{balogh2009majority}, the authors discussed the case of $\alpha=\frac{d(v)}{2}$ on hypercubes (such that $d(v)$ is the degree of vertex $v$). Flocchini et al. \cite{flocchini2003time} also studied the minimum number of blue vertices (in the initial generation) which can finally result in a completely blue generation and the necessary time for this transition to happen, especially on planar graphs, rings, and butterflies. Moreover, Mitsche and et al. \cite{mitsche2015strong} considered $\alpha=(d(v)+\alpha')/2$; they proved for any integer $\alpha'$, there exists a family of regular graphs such that with high probability all vertices become blue at the end. Recently, Koch and Lengler \cite{koch2016bootstrap} mathematically analyzed the role of geometry on bootstrap percolation for geometric scale-free networks.

The second one is the \textit{$\alpha$-threshold model} in which, at each step, a vertex becomes blue if at least $\alpha$ of its neighbors are blue, otherwise it becomes red. For instance, Schonmann \cite{schonmann1990finite} considered the state of $\alpha=\frac{d(v)}{2}$ (tie is in the favor of blue), and he showed that for any initial density of blue vertices in a torus\footnote{For a formal definition of torus and grid see Definitions \ref{definition 1} and \ref{definition 5}}, the probability of final complete occupancy by blue converges to 1 as the torus grows. Fazli et al. \cite{fazli2014non} also discussed the same model while it seems they were not aware that this model was presented (probably for the first time) in \cite{schonmann1990finite}. They presented some thresholds regarding the minimum-cardinality of an initial set of blue nodes which would eventually converge to the steady state where all nodes are blue. In addition, Moore \cite{moore1997majority} surprisingly showed that in $d$-dimensional grid for $d\geq 3$, this model can simulate boolean circuits of $AND$ and $OR$ gates.

Another model is the random \textit{$\alpha$-threshold model} such that a vertex takes the value that the $\alpha$-threshold model would give with probability $1-p$ and its complement with probability $p$. For instance, Balister et al. \cite{balister2010random} considered the case of $\alpha=\frac{d(v)}{2}$ on  2-dimensional grids. They showed that if $p$ is sufficiently small, then the process spends almost half of its time in each of two generations, all vertices blue or all red.  

The (Biased) Majority model is a subcategory of $\alpha$-threshold model, which was introduced by Gray \cite{gray1987behavior}. Majority and biased majority models are often called the(discrete time synchronous) majority-vote model in the literature. Most of prior research regarding (biased) majority model \cite{molofsky1999local,oliveira2009some,de1992isotropic,shao2009dynamic} are by physicists, and they mostly do computer simulations (i.e., Monte-Carlo methods). Specifically, their computer simulations show majority model allows stable coexistence of two colors by forming clusters of vertices holding the same color in a 2-dimensional torus. Actually, their experimental results show a phase transition behavior characterized by a large connected component of vertices holding the same color appearing when the concentrations of vertices holding the same color is above a certain threshold. However, there are some rigorous mathematical results; Poljak and Turzik \cite{poljak1986pre} presented an upper bound on the consensus time of (biased) majority model on a graph $G=(V,E)$. For instance, their results imply that for a 2-dimensional torus $G=(V,E)$, $\mathcal{O}(|V|)$ number of steps is sufficient to stabilize. Furthermore, Frischknecht, Keller, and Wattenhofer proved there exist graphs $G=(V,E)$ which need $\Omega(\frac{|V|^2}{(\log|V|)^2})$ steps to stabilize for some initial colorings in majority model. Recently, G\"artner and Zehmakan \cite{gartner2017majority} studied the behavior of the majority model on the random $d$-regular graph $\mathbb{G}_{n,d}$. It is shown that in $\mathbb{G}_{n,d}$ by starting from the initial density $p_b\leq 1/2-\epsilon$ for $\epsilon>0$, the process reaches red monochromatic configuration in $\mathcal{O}(\log_d\log n)$ steps with high probability, provided that $d\geq c/\epsilon^2$ for a suitable constant $c$.  
  
Different versions of majority updating rule have been discussed in different literatures and from various aspects for diverse aims, but as mentioned, there are just few concrete mathematical results regarding the most natural versions (majority and biased majority models). In the present paper, we study some essential and interesting aspects of the behavior of these two models and present some strong results regarding their stability, periodicity, and consensus time. As a main result we prove both majority and biased majority cellular automata exhibit a threshold behavior with two phase transitions. More precisely, we prove for a two-dimensional torus $T_{n,n}$, there are two thresholds $0\leq p_1, p_2\leq 1$ such that $p_b \ll p_1$, $p_1 \ll p_b \ll p_2$, and $p_2 \ll p_b$ result in r-monochromatic configuration, stable coexistence of both colors, and b-monochromatic configuration, respectively in $\mathcal{O}(n^2)$ number of steps.

\subsection{One-dimensional Majority Cellular Automata}
\label{One-dimensional}
  
Before going through (biased) majority model in 2-dimensional torus, we discuss the case of one-dimensional majority cellular automaton (a cycle in majority model) which might help the reader to have a better primary intuition of the techniques and the results that are presented in the rest of the paper.
  
\begin{theorem}
\label{theorem 5}
In majority model and a cycle $C_n=(V,E)$, if $p_b,p_r \gg n^{-\frac{1}{2}}$, with high probability the process reaches a bichromatic configuration at the end, but $p_b \ll n^{-\frac{1}{2}}$ results in a red monochromatic generation in at most $\frac{n}{2}$ steps.
\end{theorem}
 
\textbf{Proof:} Consider a maximum matching $M$ which divides $V$ into $\lfloor\frac{n}{2}\rfloor$ pairs $m_i$ for $1 \leq i \leq \lfloor \frac{n}{2}\rfloor$ (for $n$ odd, one vertex remains). We say a pair $m_i$ is red (blue) in generation $g$, if both of its vertices are red (blue). It is easy to see that a blue (red) pair stays blue (red) in all next generations. 

Let Bernoulli random variable $x_i$ for $1 \leq i \leq \lfloor \frac{n}{2}\rfloor$ be $1$ if and only if pair $m_i$ is blue in initial generation $g_0$. Therefore, $Pr[x_i=1]=p_{b}^{2}$. If we assume $X=\sum_{i=1}^{\lfloor\frac{n}{2}\rfloor}x_i$, then by considering $p_b \gg n^{-\frac{1}{2}}$ and $1-x \leq e^{-x}$ for $0<x<1$:
\[
Pr[X=0] \leq (1-p_{b}^{2})^{\lfloor\frac{n}{2}\rfloor} = e^{-\omega(1)}=o(1).
\]
Therefore, with high probability there exists a blue pair and also a red pair in $g_0$ (with a similar argument). 

At each step, a red (blue) path (with more than one vertex) extends its size at least by two unless it is adjacent to a red (or blue) path of size more than one. Actually, these red and blue paths grow constantly until they meet each other. Therefore, after at most $\frac{n}{2}$ steps, majority model reaches a stable generation including red and blue paths.

On the other hand, if $p_b \ll n^{-\frac{1}{2}}$, one can prove that with high probability there is no blue pair in $g_0$, but there is at least a red pair. Since there is no blue pair, this red pair extends its size at least by two at each step. Actually, each red path (with more than one vertex) stays red forever and extends its size at least by two at each step, and in an alternating path (both end points are blue and it contains no two consecutive blue (red) vertices) all internal vertices switch from blue to red (or from red to blue) and two end points get red in each step. Therefore after at most $\frac{n}{2}$ steps, it gets monochromatic by red because red (alternating) paths grow (shrink) constantly and no blue pair is created.  $\Box$ 
      
\section{(Biased) Majority Model}
\label{Not-Monochromatic Periodicity in Color War}
In this section, we first introduce two basic concepts of robust set and eternal set. Then, by using these concepts, we present sufficient conditions which guarantee the survival of a color in all upcoming generations in (biased) majority model on a graph, depending on the graph structure and the concentration of the color in the initial configuration. Specifically, we will exploit these results in Sections \ref{majority} and \ref{biased majority} to prove the threshold behavior of (biased) majority cellular automata. Furthermore, in Section \ref{Consensus Time} we discuss the number of steps which a graph $G=(V,E)$ needs to stabilize, consensus time.

\subsection{Eternal and Robust Sets}
\label{Not-monochromatic stability}
    
Recall that, if generation $g$ is a constant function, $g$ is called a \emph{monochromatic} generation, and a set $S\subseteq V$ is a $c$-community in  generation $g$ for color $c$ if $g|_S=c$. We are interested in sets of vertices that guarantee the survival of a color $c$ forever if they create a $c$-community. More specifically, we are also interested in sets of vertices which will keep a common color forever when they create a community, regardless of the colors of the other vertices.
 \begin{definition}
Let $S\subseteq V$ in a graph $G=(V,E)$. $S$ is called $c$-eternal for color $c\in \{b,r\}$ in (biased) majority model whenever the following holds: if $S$ forms a $c$-community in some generation $g_i$ for $i\geq 0$, then for all generations $g_j$ where $j\geq i$, $\exists v \in V$ such that $g_j(v)=c$.
\end{definition}
 \begin{definition}
 \label{definition 2}
In a graph $G=(V,E)$, a set $S \subseteq V$ is $c$-robust for color $c\in \{b,r\}$ in (biased) majority model whenever the following holds: if $S$ forms a $c$-community in some generation $g_i$ for $i\geq 0$, then $g_j|_S = g_i|_S=c$ for all generations $g_j$ for $j\geq i$.
\end{definition}
It follows that once a $c$-robust set forms a $c$-community, it will remain a $c$-community forever and if a $c$-eternal set creates a $c$-community once, color $c$ will survive forever. Therefore, in a graph $G=(V,E)$ a $c$-robust set is also a $c$-eternal set, but a $c$-eternal set is not necessarily a $c$-robust set. Furthermore, in majority model a set $S$ is a b-robust (b-eternal) set if and only if it is an r-robust (r-eternal) set; however, in the biased majority model an r-robust (eternal) set is a b-robust (eternal) set, but not necessarily the other way around. As a simple example, the reader might check that in a star graph $S_n$, i.e. a tree with one internal node and $n$ leaves, the internal node is the smallest eternal set (both b-eternal and r-eternal) in the majority model. What is the size of the smallest robust set in this case? 
\begin{definition}
\label{definition 3}
A blue (red) robust set in a generation $g$ is a b-robust (r-robust) set which is a b-community (r-community) in $g$.
\end{definition}
Similarly, A blue (red) eternal set in a generation $g$ is a b-eternal (r-eternal) set which is a b-community (r-community) in $g$.

Now, we discuss two theorems which present sufficient conditions, on the structure of a graph $G=(V,E)$ and the initial generation, which guarantee the survival of a color $c$ forever in (biased) majority model (without loss of generality we consider blue as color $c$).

Let for a graph $G=(V=\{v_i: 1\leq i \leq n\},E)$, $V_{j}' \subseteq V$ for $1\leq j\leq k$ be $k$ b-eternal sets; then, we define $s:=\max_{1\leq j\leq k}(|V_{j}'|)$ and $a:=\max_{1\leq i\leq n}a_i$ such that $\forall 1\leq i\leq n$ $a_i:=|\{V_{j}^{'}: v_i\in V_{j}^{'}, 1\leq j\leq k\}|$. Theorem \ref{theorem 15} says if there are $k=\omega(p_{b}^{-s})$ disjoint b-eternal sets in a graph $G=(V,E)$, then with high probability in (biased) majority model color $b$ will survive forever. On the other hand, Theorem \ref{theorem 16} explains if there are $k=\omega(\sqrt{n}p_{b}^{-s})$ (not necessarily disjoint) b-eternal sets in graph $G=(V,E)$ and $a$ is a constant, then color $b$ survives with high probability.                           
\begin{theorem}
\label{theorem 15}
For a graph $G=(V,E)$ and (biased) majority model, if $V_{j}'$ for $1\leq j\leq k$ are disjoint b-eternal sets, then $Pr[$r-monochromatic$]\leq \exp(-kp_{b}^{s})$. 
\end{theorem}
\textbf{Proof:} We define $k$ random variables $x_j$ such that for $1\leq j\leq k$:

\[ 
x_j = \left\{ 
\begin{array}{rl} 
1 &  if g_0|_{V_{j}^{'}}=b \\
0 & otherwise
\end{array} \right. 
\]
 where $Pr[x_j=1]=p_{b}^{|V_{j}'|}$  for $1 \leq j \leq k$. Therefore, if $X:=\sum_{j=1}^{k}x_j\geq 1$, then (biased) majority model does not get r-monochromatic because there is at least a blue eternal set. We show $Pr[X=0]\leq \exp(-kp_{b}^{s})$ which implies $Pr[$r-monochromatic$]\leq \exp(-kp_{b}^{s})$. 
 \\
$X$ is the summation of $k$ independent Bernoulli random variables, then:
 \[
Pr[X=0]\leq (1-p_{b}^{s})^{k} \leq e^{-kp_{b}^{s}} .\quad \Box
\]
\begin{theorem}
\label{theorem 16}
In a graph $G(V=\{v_i: 1\leq i\leq n\},E)$ and (biased) majority model, if $V_j^{'}$ $\forall 1\leq j\leq k$ are b-eternal sets (not necessarily disjoint), then $Pr[$r-monochromatic$]\leq \exp(-k^2p_{b}^{2s}/2\sum_{i=1}^{n} a_{i}^{2})$. 
\end{theorem} 
\textbf{Proof:} Let random variable $X$ denote the number of sets $V_{j}'$ for $1\leq j\leq k$ which create a b-community in $g_0$ i.e. $g_0|_{V_{j}'}=b$. If we prove $Pr[X=0]\leq \exp(-k^2p_{b}^{2s}/2\sum_{i=1}^{n}c_{i}^{2})$, then $Pr[$r-monochromatic$]\leq \exp(-k^2p_{b}^{2s}/2\sum_{i=1}^{n}a_{i}^{2})$. 

If we are given $n$ discrete probability spaces $(\Omega_i,Pr_i)$ $1\leq i\leq n$, then their product is defined to be the probability space over the base set $\Omega:=\Omega_1 \times$...$\times \Omega_n$ with the probability function
\[
Pr[(\omega_1,...,\omega_n)]=\prod_{i=1}^{n}Pr_i[\omega_i]
\] 
where $\omega_i\in \Omega_i$. Now, we have random variable $X:\Omega \rightarrow \mathbb{R}$ so that $(\Omega,Pr)$ is the product of $n$ discrete probability spaces which correspond to independent random coloring of all vertices in $g_0$.

We say that the effect of the $i$-th coordinate is at most $a_i$ if for all $\omega, \omega'\in \Omega$ which differ in the $i$-h coordinate we have $|X(\omega)-X(\omega')|\leq a_i$. $a_i$ is actually equal to the number of b-eternal sets which contain vertex $v_i$ i.e $a_i:=|\{V_{j}^{'}: v_i\in V_{j}^{'}, 1\leq j\leq k\}|$.
\\
Now, by utilizing Azuma's Inequality \cite{feller1968introduction}, we have:
\[
Pr[X\leq E[X]-t]\leq e^{-\frac{t^2}{2\sum_{i=1}^{n}a_{i}^{2}}}, \quad \forall t>0
\]
\\
We have $E[X]\geq kp_{b}^{s}$, therefore:
\[
Pr[X=0]\leq e^{-\frac{k^2p_{b}^{2s}}{2\sum_{i=1}^{n}a_{i}^{2}}}. \quad \Box
\]
 
\begin{corollary}
\label{corollary 1}
If $k=\omega(\sqrt{n}p_{b}^{-s})$ and $a$ is constant, then $Pr[$r-monochromatic$]=o(1)$.
\end{corollary}

\subsection{Periodicity and Consensus Time}
\label{Consensus Time}
For a graph $G=(V,E)$ and (biased) majority model, the number of possible generations is $2^{|V|}$, and (biased) majority model is a deterministic process; therefore, the process always reaches a cycle of generations after a finite number of steps and stays there forever, but there are two natural questions which arise. What is the length of the cycle and how long does it take to reach it?

Goles and Olivos \cite{goles1981comportement} and independently Poljak and Sura \cite{poljak1983periodical} proved that a large class of majority-based models, including (biased) majority model, always reach a cycle of period one or two. More precisely, they consider a set $V$ of individuals such that every $v \in V$ has an initial color from set $\{0,1, \cdots,l\}$ for $l\in \mathbb{N}$ i.e. there is a function $f_0:V\rightarrow\{0,\dots,l\}$. Furthermore, the function $\omega(u,v)$ for $u,v \in V$ measure the influence of $u$ on $v$ and it is symmetric which means $\forall u,v \in V$ $\omega(u,v)=\omega(v,u)$. Now, consider a system $(V,\omega,f_0)$ which evolves over time so that for every $t\geq 0$, the function $f_{t+1}:V\rightarrow\{0,\dots,l\}$ which maps a member to its color at time $t+1$ is defined as follows for $u\in V$ ($u$ adopts the most frequent color in its neighborhood and in case of a tie, it chooses the largest one.):
\[
f_{t+1}(u)=\max\{i:\forall j \sum_{f_{t}(v)=i, v\in V}\omega(u,v)\geq\sum_{f_{t}(v)=j, v\in V}\omega(u,v)\}.
\]
Now, assume the period of a system is defined as minimum $t>0$ so that $f_{i+t}=f_{i}$ for some $i$. Then, Goles and Olivos \cite{goles1981comportement} and Poljak and Sura \cite{poljak1983periodical} present Theorem \ref{theorem 6}.
\begin{theorem} \cite{goles1981comportement}
\label{theorem 6}
A system $(V,\omega,f_0)$ with symmetric $\omega$ following 
\[
f_{t+1}(u)=\max\{i:\forall j \sum_{f_{t}(v)=i, v\in V}\omega(u,v)\geq\sum_{f_{t}(v)=j, v\in V}\omega(u,v)\}
\] 
always reaches a cycle of period one or two.
\end{theorem}
\begin{corollary}
\label{theorem 40}
In (biased) majority model, an arbitrary graph $G=(V,E)$ (from any initial coloring) always reaches a cycle of generations of length one or two.
\end{corollary} 
\textbf{Proof:} Let $l=1$ and two colors $0$ (red) and $1$ (blue). The initial generation $g_0$ corresponds to $f_0$ and we consider symmetric function $\omega$ as follow
\[
\omega(u,v)= \left\{\begin{array}{lll}1, &\mbox{if $u\in N(u)$}, \\
\frac{1}{2},&\mbox{if u=v}, \\
0,&\mbox{otherwise}
\end{array}\right.
\]
One can easily see that in this case the model is equivalent to the majority model. Therefore, the proposition is true in this case. For the case of biased majority model is sufficient to change the state of $u=v$ from $1/2$ to $0$. $\Box$

As mentioned, we are also interested in the consensus time of the process, and it is trivial that it is at most $2^{|V|}$. However, Poljak and Turzik \cite{poljak1986pre} proved a very strong proposition (see Theorem \ref{theorem 1}), in the literature of cyclically monotonous mappings and symmetric matrices which provides a tight upper bound on the consensus time of (biased) majority model. More precisely, they present an upper bound on the pre-period of mappings of form $g(x)=f(Ax)$ where $A$ is a linear mapping given by a symmetric matrix of size $n \times n$ and the vector $x\in\{-1,+1\}^n$ (for more details see Theorem \ref{theorem 1}). Furthermore, pre-period means the maximal $k$ such that all $g(x)$, $g^2(x)$, $\cdots$, $g^k(x)$ are distinct which is equivalent to the consensus time of the process in our terminology.
\begin{theorem} \cite{poljak1986pre}
\label{theorem 1}
Let $f:\mathbb{Z}^n\rightarrow\{-1,+1\}^n$ be defined by $f(x^1, \dots, x^n)=(y^1,\dots,y^n)$ where $y^i=1$ if $x^i\geq0$ and $y^i=-1$ if $x^i<0$. Let $A=(a_{ij})$ be a symmetric matrix with integral entities, then the pre-period of $fA:\{-1,+1\}^n\rightarrow\{-1,+1\}^n$ is at most $\frac{1}{2}(\sum_{i,j}|a_{ij}|+3s-n)$ where $s=|\{i:\sum_{j=1}^{n}a_{ij}\text{is even}\}|$.
\end{theorem} 
\begin{corollary}
\label{corollary 7}
In a graph $G=(V,E)$ for majority model and biased majority model, the consensus time is at most $|E|$ and $|E|+|V|$, respectively.
\end{corollary}
\textbf{Proof:} Set $n=|V|$. Let $v_1, \dots, v_n$ be an arbitrary ordering of vertices in $V$ and let $g_0(v_i) \in\{b,r\}$ denote the initial coloring of the vertices. The entries of the matrix $A=(a_{ij})$ of size $n\times n$ are defined as follows for $1 \leq i \ne j \leq n$:
\[
a_{ij}= \left\{\begin{array}{lll}1, &\mbox{if $v_i\in N(v_j)$}, \\
0,&\mbox{otherwise}
\end{array}\right.
\] 
and for $1 \leq i \leq n$ as follows:
\[
a_{ii}= \left\{\begin{array}{lll}1, &\mbox{if $d(v_i)$ $\mbox{is even}$}, \\
0,&\mbox{otherwise}
\end{array}\right..
\]
$A$ is a symmetric matrix since $v_i \in N(v_j)$ if and only if $v_j \in N(v_i)$.

Set $x^i=-1$ if $g_{0}(v_i)=r$ and $x^i=+1$ if $g_0(v_i)=b$. For a vertex $v_i \in V$ with $d(v_i)$ odd, there is always a color that occurs most frequently in the neighborhood since the number of neighbors is odd. Note that $(Ax)^i<0$ if the majority of the neighbors of $v_i$ is red and $(Ax)^i>0$ otherwise. For a vertex $v_i \in V$ with $d(v_i)$ even, the rules of majority model state that in the following generation $v_i$ takes the color that occurs most frequently in its neighborhood and in case of a tie, it conserves its color. In other words, $v_i$ takes the color that occurs most frequently in $\hat{N}(v_i)$. Note that $(Ax)^i<0$ if the majority of the vertices in $\hat{N}(v_i)$ is red and $(Ax)^i>0$ otherwise. This shows that applying the mapping $fA$ to the vector $x\in\{-1,+1\}^n$ corresponds to one step in majority model on the graph $G=(V,E)$.

Furthermore, $s=|\{i:\sum_{j=1}^{n}a_{ij} \text{is even}\}|=0$ since every row of $A$ contains an odd number of $1$’s and every other entry is $0$ by construction. Theorem \ref{theorem 1} implies that the consensus time is at most $\frac{1}{2}(\sum_{i,j}|a_{ij}|-n)$. For $v_i$ with $d(v_i)$
even, it holds that $\sum_{j}|a_{ij}|=d(v_i)+1$ and for $v_i$ with $d(v_i)$ odd, it holds that $\sum_{j}|a_{ij}|=d(v_i)$. Therefore, the consensus time is at most $\frac{1}{2}(\sum_{v\in V}d(v)+r_e+3s-n)$ which is equal to $\frac{1}{2}(\sum_{v\in V}d(v)-r_o)$ because $s=0$ (where $r_e$ and $r_o$ are equal to the number of vertices with even degree and odd degree, respectively) which is obviously smaller than $\frac{1}{2}(\sum_{v\in V}d(v))$ which is equal to $|E|$. 

The case of biased majority model also can be proved very similarly by defining  for $1 \leq i,j \leq n$:
\[
a_{ij}= \left\{\begin{array}{lll}1, &\mbox{if $v_i\in N(v_j)$}, \\
0,&\mbox{otherwise}
\end{array}\right.. \Box
\] 
               
\section{(Biased) Majority Model in Grid and Torus}
\label{Monochromatic Stability in Torus and Grid} 
In this section, first some primary definitions concerning torus and grid are presented. Then, it is proved that majority cellular automata and biased majority cellular automata show a threshold behavior respectively in Section \ref{majority} and Section \ref{biased majority}. Furthermore, we also discuss the consensus time and the periodicity of both automata by exploiting the aforementioned results in Sections \ref{Not-Monochromatic Periodicity in Color War}.

\subsection{Preliminaries}
\begin{figure}[h]
\begin{center}
\includegraphics[width=0.5\textwidth]{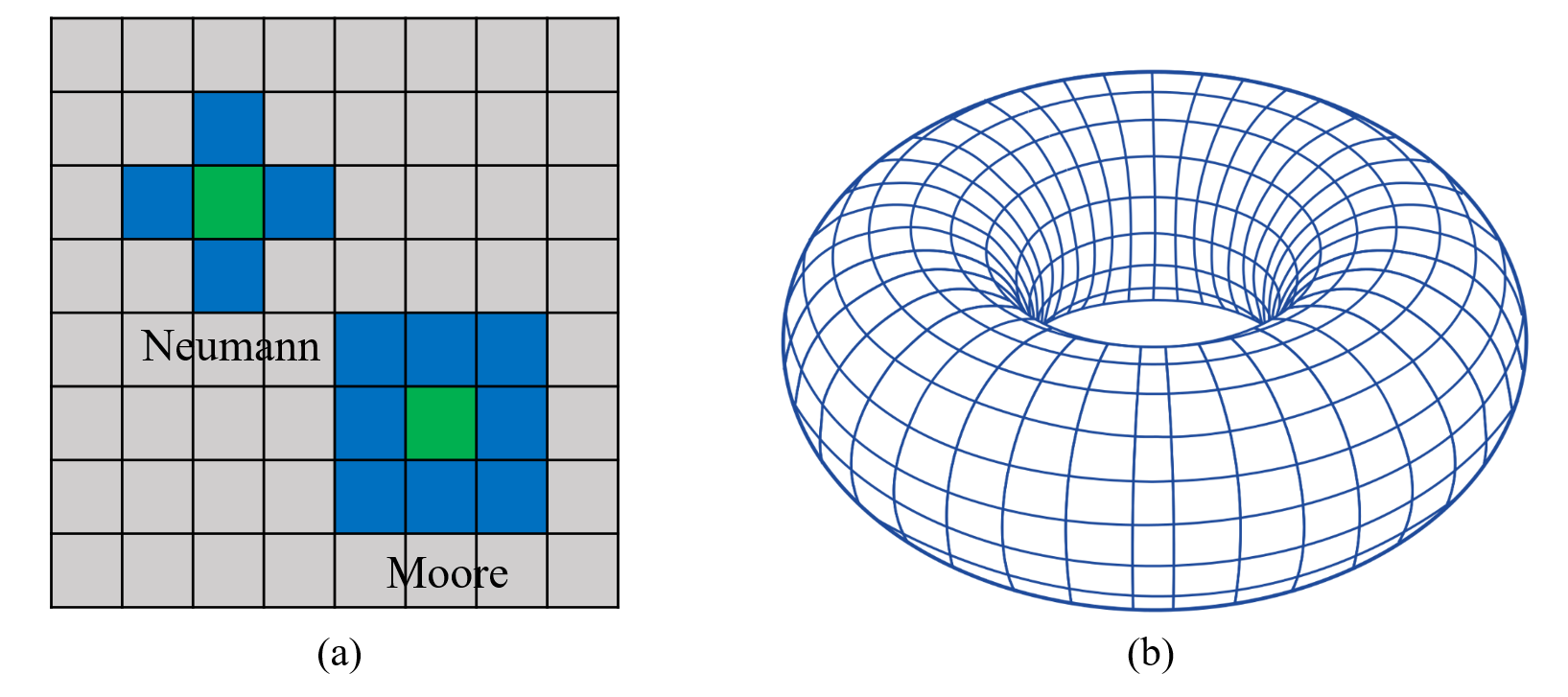}
\caption{(a) Neumann and Moore neighborhoods in the grid $G_{8 \times 8}$ (b) a torus\label{fig 11}}
\end{center}
\end{figure}

 \begin{definition}
 \label{definition 1}
 The grid $G_{n,n}$ is the graph $G=(V,E)$ such that $V=\{(i,j): 0\leq i,j\leq n-1\}$ and $E=\{((i,j),(i',j')): (i=i' \wedge |j-j'|=1) \vee (|i-i'|=1 \wedge j=j')\}$.
 \end{definition}
  \begin{definition}
  \label{definition 5}
  The torus $T_{n,n}$ is the graph $G=(V,E)$ such that $V=\{(i,j): 0\leq i,j\leq n-1\}$ and $E=\{((i,j),(i',j')): (i=i' \wedge (|j-j'|=1 \vee |j-j'|=n-1)) \vee ((|i-i'|=1 \vee |i-i'|=n-1) \wedge j=j')\}$.
  \end{definition}
   A torus $T_{n,n}$ is a wrap-around version of the grid $G_{n,n}$ which can be visualized as taping the left and right edges of the rectangle to form a tube, then taping the top and bottom edges of the tube to form a torus (See Figure \ref{fig 11} (b)).
   \\
   The aforementioned definitions of grid and torus follow a neighborhood model which is called Neumann neighborhood or 4-neighborhoods (see Figure \ref{fig 11} (a)). On the other hand, there is another common neighborhood model which is called Moore model (see Figure \ref{fig 11} (a)), and in a torus or grid (by skipping the borders), each cell instead of four neighbors has eight neighbors. More accurately, the grid $G_{n,n}$ with Moore neighborhood is the graph $G=(V,E)$ such that $V=\{(i,j): 0\leq i,j\leq n-1\}$ and $E=\{((i,j),(i',j')): (|i-i'|\leq 1 \wedge |j-j'| \leq 1 \wedge |i-i'|+|j-j'|\ne 0)$.
  \begin{remark}
  We sometimes use the term of \emph{cell} instead of vertex in grids and tori.
  \end{remark}
\begin{definition}
$\forall 0\leq i\leq n-1$, column $c_i:=\{(i,j): 0\leq j\leq n-1\}$ and $\forall 0\leq j\leq n-1$ row $r_j:=\{(i,j): 0\leq i\leq n-1\}$.
\end{definition}

\subsection{Majority Cellular Automata}
\label{majority}
In this section, we prove that $n^{-\frac{1}{2}}$ is a threshold for getting monochromatic or bichromatic in majority cellular automata with Neumann neighborhood. More precisely, $p_b,p_r \gg n^{-\frac{1}{2}}$ results in a cycle of bichromatic generations, but $p_b \ll n^{-\frac{1}{2}}$ outputs a red monochromatic generation with high probability. For proving the first part, we exploit the concept of robustness and show that if $p_b,p_r \gg n^{-\frac{1}{2}}$, there exists a blue robust and a red robust set in $g_0$ with high probability which guarantee reaching a cycle of bichromatic generations. On the other hand, for proving the second part we consider a constant number of initial generations instead of just considering initial generation $g_0$, and interestingly, this technique helps us to prove that $p_b \ll n^{-\frac{1}{2}}$ results in a red monochromatic generation with high probability. Furthermore, it is shown the aforementioned threshold property works also for Moore neighborhood by considering $n^{-\frac{1}{6}}$ instead of $n^{-\frac{1}{2}}$. 
\begin{remark}
As we know, in the case of majority model, there is no difference between red and blue in the sense of updating rule; then, in this section we simply utilize the term of robust set instead of b-robust set and r-robust set and without loss of generality we also assume $p_b \leq p_r$. 
\end{remark}

\begin{figure}[h]
\begin{center}
\includegraphics[width=0.4\textwidth]{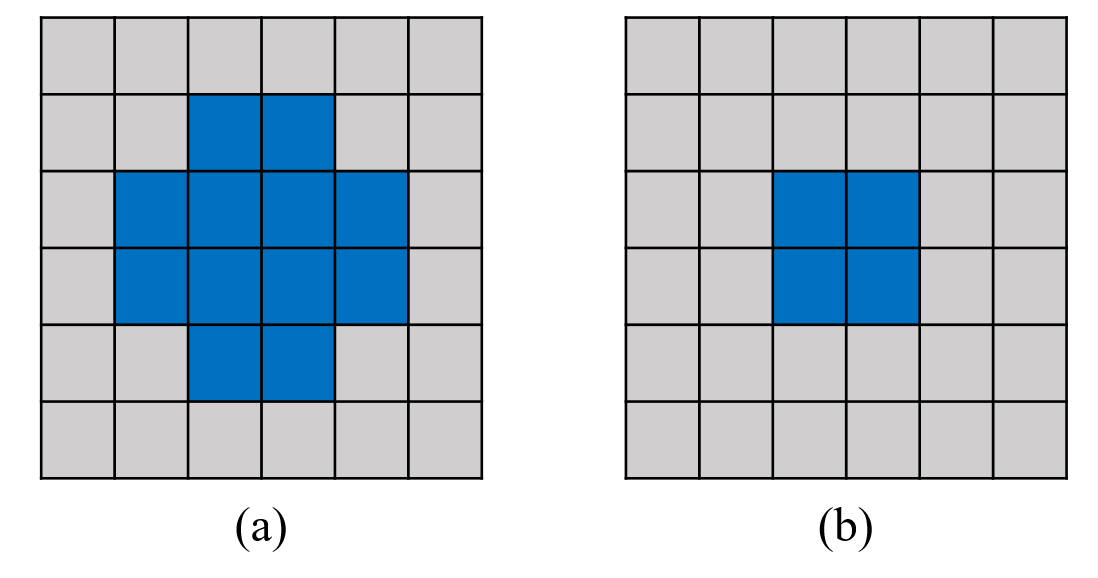}
\caption{The smallest robust set in the majority model: (a) for a torus with Moore neighborhood (b) Neumann neighborhood \label{fig 9}}
\end{center}
\end{figure}

In the torus $T_{n,n}$ with Neumann neighborhood (Moore neighborhood) there is a robust set of size $4$ ($12$) as shown in Figure \ref{fig 9} (assume $n$ is large enough, say $n>4$ ($n>12$) in the case of Neumann (Moore) neighborhood). Actually, in the proof of Theorem \ref{theorem 30} (Theorem \ref{theorem 29}), we show that in the torus $T_{n,n}$ with Neumann (Moore) neighborhood if in a generation $g$, there are at most $3$ ($11$) blue vertices, the process reaches an r-monochromatic generation in constant number of steps; therefore, the size of the smallest robust set in the case of Neumann and Moore neighborhood is $4$ and $12$, respectively. As we will discuss, the size of the smallest robust set plays a critical role in the threshold behavior of majority cellular automaton.

For the torus $T_{n,n}$ with Neumann neighborhood, a cluster is a connected subgraph by considering the same vertex set but Moore neighborhood instead of Neumann neighborhood. Therefore, in a torus with Neumann neighborhood, a cluster is not necessarily a connected component. 
                                                                                   
\begin{definition}
For the torus $T_{n,n}$ with Neumann neighborhood, $L_{j,b}$ ($L_{j,r}$) is the size of the largest blue (red) cluster in generation $g_j$ for $j\geq 0$.
\end{definition}                              
\begin{lemma}
 \label{lemma 5}
In majority model, the torus $T_{n,n}$ with Neumann neighborhood and for a generation $g_{j}$ $j\geq 0$, if $L_{j,b}\leq 3$, then $L_{j+2,b}=0$.
\end{lemma}
\textbf{Proof:} Let $L_{j}:=L_{j,b}$ and $L_{j+1}:=L_{j+1,b}$. First we prove if $L_{j}=1$, then $L_{j+1}=0$. To prove, assume $L_j=1$ can result in $L_{j+1}=1$ i.e there exists a cell $c_1$ such that $g_{j+1}(c_1)=b$. We show $L_{j+1}=1$ contradicts $L_j=1$. $g_{j}(c_1)=r$ because a blue cell for staying blue in $g_{j+1}$ needs at least two blue cells in its neighborhood in $g_j$ which contradicts $L_{j}=1$. $g_{j}(c_1)=r$ implies that $|N_{b}^{g_{j}}(c_1)|\geq 3$, and $|N_{b}^{g_{j}}(c_1)|\geq 3$ results in the existence of at least a blue cluster of size $2$ in $c_1$'s neighborhood in $g_{j}$ which contradicts $L_{j}=1$.
 
\begin{figure}[h]
\begin{center}
\includegraphics[width=0.5\textwidth]{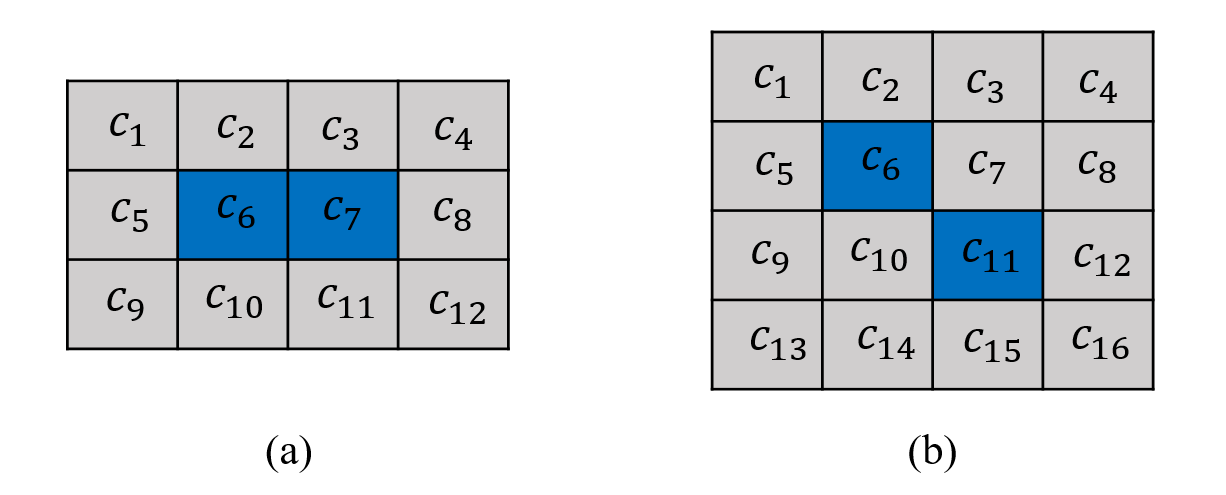}
\caption{Two possible structures for a cluster of size $2$ in a torus\label{fig 14}}
\end{center}
\end{figure}

Now, it is proved that if $L_{j}=2$ or $3$, then $L_{j+1}\leq 1$ i.e. blue clusters of size $3$ or smaller cannot create a blue cluster of size larger than $1$. It is enough to prove $L_{j+1}=2$ implies $L_{j}\geq 4$. Assume there exists a blue cluster $S$ of size $2$ in generation $g_{j+1}$. $S$ can have two different structures which are shown in Figure \ref{fig 14} (notice a torus is symmetric).
\\
For the first structure $S=\{c_6,c_7\}$ (see Figure \ref{fig 14} (a)), we have three states:
\\
\textbf{(i)} if $g_{j}(c_6)=g_{j}(c_7)=b$, then $c_6$ for staying blue in $g_{j+1}$ needs at least a blue cell among $\{c_2,c_5,c_{10}\}$ in $g_{j}$ and $c_7$ needs at least a blue cell among $\{c_3,c_8,c_{11}\}$ which result in $L_{j}\geq 4$.
\\
\textbf{(ii)} if $g_j(c_6)=b$ and $g_j(c_7)=r$ (or similarly $g_j(c_6)=r$ and $g_j(c_7)=b$), then $c_{7}$ for getting blue in $g_{j+1}$ needs at least two blue cells among $\{c_3,c_8,c_{11}\}$ in $g_j$, and $c_6$ needs at least one blue cell among $\{c_2,c_5,c_{10}\}$ which imply $L_{j}\geq 4$.
\\
\textbf{(iii)} if $g_j(c_6)=g_{j}(c_7)=r$, then $g_{j+1}|_{\{c_6,c_7\}}=b$ implies $g_j|_{\{c_2,c_3,c_5,c_8,c_{10},c_{11}\}}=b$ which means $L_{j}\geq 4$.
\\
For the second structure (see Figure \ref{fig 14}), also there are three possibilities:
\\
\textbf{(i)} if $g_{j}(c_6)=g_j(c_{11})=b$, then $c_6$, for staying blue in $g_{j+1}$, needs at least two blue cells among $\{c_2,c_5,c_7,c_{10}\}$ in $g_j$ which implies $L_{j}\geq 4$.
\\
\textbf{(ii)} if $g_j(c_6)=b$ and $g_j(c_{11})=r$ (or similarly $g_j(c_6)=r$ and $g_j(c_{11})=b$), then $c_{11}$, for getting blue in $g_{j+1}$, needs at least three blue cells among its neighbors ($\{c_7,c_{10},c_{12},c_{15}\}$) in $g_j$ which implies $L_j\geq 4$ again.
\\
\textbf{(iii)} if $g_j(c_6)=g_{j}(c_{11})=r$, then $g_{j+1}|_{\{c_6,c_{11}\}}=b$ implies that three cells in $\{c_2,c_5,c_7,c_{10}\}$ and three cells in $\{c_7,c_{10},c_{12},c_{15}\}$ are blue in $g_j$ which mean that $L_j \geq 4$.

Therefore, $L_{j+1}=2$ implies $L_j \geq 4$ which means $L_j\leq 3$ results in $L_{j+1}\leq 1$. Furthermore, we proved $L_j=1$ outputs $L_{j+1}=0$. Then, $L_j\leq 3$ provides $L_{j+2}=0$. $\Box$                                                      
\begin{corollary}
\label{corollary 6}
In the torus $T_{n,n}=(V,E)$ with Neumann neighborhoods and a generation $g_{j}$, if $L_{j,b}\leq 3$, then generation $g_{j+2}$ is red monochromatic.
\end{corollary}
\begin{theorem}
\label{theorem 30}
In the majority model and the torus $T_{n,n}=(V,E)$ with Neumann neighborhood, $p_b\ll n^{-\frac{1}{2}}$ results in a red monochromatic generation in at most $2$ steps almost surely, but $p_b\gg n^{-\frac{1}{2}}$ outputs a cycle of bichromatic generations of size one or two in $\mathcal{O}(n^2)$ steps with high probability. 
\end{theorem} 
\textbf{Proof:} First consider the case of $p_b\ll n^{-\frac{1}{2}}$ in torus $T_{n,n}$. Let random variable $X$ be the number of blue clusters of size $4$ in $g_0$. We claim $E[X]=o(1)$, then by Markov's Inequality \cite{feller1968introduction} with the probability of $1-o(1)$, $L_{0,b}\leq 3$. Based on Corollary \ref{corollary 6}, $g_2|_V=r$ i.e. $g_2$ is red monochromatic. Then, it is enough to show that $E[X]=o(1)$. Since the number of clusters of size $4$ in $T$ is $\Theta(n^2)$ (every vertex could be contained in at most a constant number of clusters of size $4$):

\[
E[X]= \Theta(n^2)p_{b}^{4} = o(1).
\]
Now, we discuss the state of $p_b\gg n^{-\frac{1}{2}}$. Consider $\forall 1\leq i'< \lfloor \frac{n}{2} \rfloor-1$ $\wedge$ $\forall 1\leq j'< \lfloor \frac{n}{2} \rfloor-1$ $S_{i',j'}=\{(i,j)| 2i'-1\leq i\leq 2i' \wedge 2j'-1\leq j\leq 2j'\}$ (see Figure \ref{fig 9} (b)) as $\Theta(n^2)$ disjoint robust sets. Based on Theorem \ref{theorem 15}, majority model in this state with high probability reaches a cycle of bichromatic generations. Actually, by utilizing Theorem \ref{theorem 40} and Corollary \ref{corollary 7}, we can say it reaches a cycle of bichromatic generations of size one or two in $\mathcal{O}(n^2)$ steps. $\Box$ 
                         
\begin{theorem}
\label{theorem 29}
In majority model and the torus $T_{n,n}=(V,E)$ with Moore neighborhood, $p_b\ll n^{-\frac{1}{6}}$ results in an r-monochromatic generation in constant number of steps, but $p_b\gg n^{-\frac{1}{6}}$ outputs a cycle of bichromatic generations of size one or two in $\mathcal{O}(n^2)$ steps with high probability. 
\end{theorem}
\textbf{Proof:}
First, we prove $p_b \ll n^{-\frac{1}{6}}$ outputs a red monochromatic generation in constant number of steps with high probability. Let for arbitrary $0 \leq i',j'\leq n-1$ and $1 \leq l_1,l_2 \leq n$, set $R(i',j',l_1,l_2):=\{(i \mod n,j \mod n): i'\leq i<i'+l_1 \wedge j'\leq j<j'+l_2\}$ be a rectangle of size $l_1\times l_2$ in torus $T_{n,n}$. Let random variable $X$ denote the number of squares of size $23\times 23$ (squares $R(i,j,23,23)$ for $0 \leq i,j \leq n-1$) in the torus $T_{n,n}$ which include more than $11$ blue vertices in $g_0$. Therefore:
\[
E[X] \leq \Theta(n^2)\sum_{i=12}^{23^2}{23^2 \choose i}p_{b}^{i}=\Theta(n^2)p_{b}^{12}\sum_{i=12}^{23^2}{23^2 \choose i}p_{b}^{i-12}\leq \Theta(n^2)p_{b}^{12}\sum_{i=12}^{23^2}{23^2 \choose i}=o(1).
\]
$E[X]=o(1)$ and Markov's Inequality imply that with high probability $X=0$, i.e. with high probability there is no square of size $23 \times 23$ in $T_{n,n}$ which contains more than $11$ blue cells in $g_0$.

Let for a set $S \subseteq V$, $R_S:=\{R(i,j,l_1,l_2): 0 \leq i,j \leq n-1 \wedge 1 \leq l_1,l_2 \leq n \wedge S\subseteq R(i,j,l_1,l_2)\}$, then we define that rectangle $R \subseteq V$ is the \textit{smallest covering rectangle} of $S$ if $R$ is the smallest rectangle which covers $S$ i.e. $R:=\argmin_{R'(i,j,l_1,l_2)\in R_S} |l_1 \times l_2|$. Based on the definition, the smallest covering rectangle for a set $S$ is not necessarily unique. Furthermore, for two vertices (cells) $u,v \in V$, the distance $d(u,v)$ is the size of the shortest path between $u$ and $v$ in terms of the number of edges minus one in the torus $T_{n,n}=(V,E)$ with Moore neighborhood (for instance, the distance between two neighbor vertices is zero; we also define $d(v,v)=0$ for a vertex $v$); then for two rectangles $R_1,R_2 \subset V$, $d(R_1,R_2):=\min_{u\in R_1,v\in R_2}d(u,v)$. Consider the following procedure on the torus $T_{n,n}$ with initial generation $g_0$ ($p_b \ll n^{-\frac{1}{6}}$) where $M$ is the set of the smallest covering rectangles of all connected blue components in $g_0$ (as we mentioned, the smallest covering rectangle for a set is not necessarily unique, but in this proof considering any smallest covering rectangle for a set works), and for two rectangles $R$ and $R'$, $Combine(R,R')$ denotes the smallest rectangle which covers both $R$ and $R'$.

\begin{algorithm}
    \SetKwInOut{Input}{Input}
    \SetKwInOut{Output}{Output}

    \underline{Rectangulation Procedure}\\
    \Input{set $M$ of rectangles}
       $M'=M$\;
 \While{$\exists R,R'\in M' \quad s.t. \quad d(R,R')\leq 1$}{
  $M'=M'\setminus\{R,R'\}\cup\{Combine(R,R')\}$\;
 }
      {
        return set $M'$\;
      }
    \caption{Rectangulation of blue cells\label{algorithm}}
\end{algorithm}
 
After the aforementioned procedure, for every rectangle $R \in M'$ of size $l_1 \times l_2$, we have $l_1, l_2 \leq 23$ with high probability because based on the process, $l_1>23$ or $l_2>23$ implies that there exists a rectangle of size $23 \times 23$ which includes more than $11$ blue cells in $g_0$. More precisely, one can show by induction that every rectangle in $M'$ contains at least a blue cell in every two consecutive columns and a blue cell in every two consecutive rows for rows and columns which intersect the rectangle. On the other hand, as we proved with high probability for $p_b \ll n^{-\frac{1}{6}}$, there is no rectangle of size $23 \times 23$ which contains more than $11$ blue cells. Therefore, after this process, $\forall R,R' \in M'$, the number of blue cells in $R$ is at most $11$ with high probability and $d(R,R')\geq 2$ i.e. $M'$ covers blue cells in $g_0$ with rectangles which have at most $11$ blue cells inside and the shortest distance between each pair of rectangles is more than $1$.

Since the distance between each two rectangles is at least two, all cells out of these rectangles stay red in all upcoming generations. Therefore, if we show that $11$ blue cells in a rectangle disappear (rectangle gets completely red) after a constant number of steps, then the proof is complete and we can say $p_b \ll n^{-\frac{1}{6}}$ results in a red monochromatic generation after a constant number of steps with high probability.

Now, we demonstrate that $11$ blue cells in a rectangle surrounded by red cells disappear in a constant number of steps. More precisely, we prove $i$ blue cells are reduced to at most $i-1$ blue cells in one or two steps for $5 \leq i \leq 11$, and $i$ blue cells disappear in one step for $i=1,2,3,4$. Therefore, $11$ blue cells disappear in a constant number of steps. 

 Consider a rectangle $R \in M'$ and assume $S$ is the set of blue cells in $R$ in generation $g_0$ and $S'$ is the set of blue cells in $R$ in generation $g_1$ (assume $|S|=s$ and $|S'|=s'$). Firstly, $s=i$ blue cells will disappear in one step for $1\leq i\leq 4$ because $g_1(v)=b$ for a cell $v \in V$ implies $|\hat{N}_{b}^{g_0}(v)|\geq 5$. 

Define $Sh(u,v):=|\hat{N}(v)\cap\hat{N}(u)|$ for $u,v \in V$. Now, we present the following proposition which we exploit in the rest of the proof several times.

\textbf{Proposition 1}: In the torus $T_{n,n}=(V,E)$ with Moore neighborhood and for two vertices $u,v \in V$, if $g_1|_{\{v,u\}}=b$, then $|\hat{N}_{b}^{g_0}(\{u,v\})|\geq 10-Sh(u,v)$.

The proposition is true because as we know, vertex $v$ ($u$) needs at least $5$ blue cells in $\hat{N}(v)$ ($\hat{N}(u)$) in $g_0$ to get blue in $g_1$, and the number of cells which they can share is $Sh(u,v)$ which finishes the proof of Proposition 1.

One can easily check that $s=5$ blue cells can create at most $2$ blue cells in the next generation; $s'>2$ implies there exist two cells $u,v \in S'$ such that $Sh(u,v)\leq 4$ which means $s\geq 6$ because of Proposition 1.

$s=6$ blue cells can create at most $4$ blue cells in the next step. Assume they can create more then $4$ blue cells i.e. $s' \geq 5$. If $R'$ of size $l_1 \times l_2$ is the smallest covering rectangle of $S'$, then $l_1 \geq 3$ or $l_2\geq 3$ because it contains at least $5$ blue cells. Without loss of generality assume $l_1\geq 3$, then there exist $u,v \in S'$ such that $Sh(u,v)\leq 3$ (for instance, consider the leftmost and rightmost cells in $S'$) which implies $s\geq 10-3=7$ because of Proposition 1.

\begin{figure}[h]
\begin{center}
\includegraphics[width=0.25\textwidth]{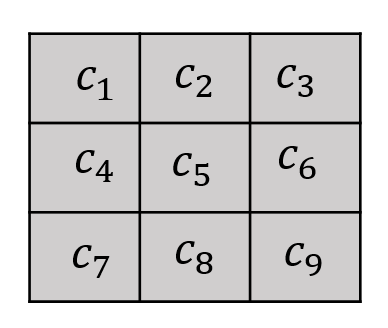}
\caption{$R'$ in the case of $l_1=l_2=3$\label{fig 13}}
\end{center}
\end{figure}

We show $7$ blue cells can create at most $6$ blue cells in the next step. Again, we use the same idea; assume set $S$ with $7$ blue cells can create $s'$ blue cells such that $s'\geq 7$, then there are two possibilities. First, if $l_1 \geq 4$ or $l_2\geq 4$ in $R'$ of size $l_1 \times l_2$ (the smallest covering rectangle of $S'$), then there exist two cells $u,v \in S'$ such that $Sh(u,v)=0$ (for instance, the leftmost (high-most) and the rightmost (low-most) cells in $S'$ in the case of $l_1\geq 4$ ($l_2\geq 4$)) which results in $s \geq 10$ by Proposition 1. Otherwise, $l_1=l_2=3$; in this case by considering Figure \ref{fig 13} as $R'$, if $\{c_1,c_9\}\subset S'$ (or similarly $\{c_3,c_7\} \subset S'$), then $s\geq 10-1=9$ by Proposition 1 in that $Sh(c_1,c_9)=1$. If $\{c_1,c_9\} \nsubseteq S'$ and $\{c_3,c_7\} \nsubseteq S'$, then $S'$ must contain exactly one cell of set $\{c_1,c_9\}$ and one cell of set $\{c_3,c_7\}$ because $s'\geq 7$. Without loss of generality, assume $S'=R'\setminus\{c_7,c_9\}$. Again, $Sh(c_4,c_3)=2$ which results in $s\geq 10-2=8$ by Proposition 1. 

In proving that $s=8$ cannot result in $s'\geq 8$, the case of $l_1\geq 4$ or $l_2\geq 4$ with the same argument for $7$ blue cells (see the previous paragraph) implies $s\geq 10$. Furthermore, if $l_1=l_2=3$, based on $s'\geq 8$, $\{c_1,c_9\}\subset S'$ or $\{c_3,c_7\} \subset S'$ (see Figure \ref{fig 13}). Since $Sh(c_1,c_9)=1$ and $Sh(c_3,c_7)=1$, $s\geq 10-1=9$ by Proposition 1.

Now, we prove if there are $9$ blue cells ($s=9$), in one step or at most two steps there exist at most $8$ blue cells. If $l_1\geq 4$ or $l_2\geq 4$, with the aforementioned argument for the cases of $7$ and $8$ blue cells, we have $s\geq 10$. Otherwise, $l_1=l_2=3$ and $S'=R'$; in this case, the number of blue cells in the second step is $5$ because all cells in the corners ($c_1,c_3,c_7,$ and $c_9$: see Figure \ref{fig 13}) become red.

We claim $10$ blue cells also cannot create more than $9$ blue cells in the next step or at most two next steps. Assume a set $S$ of $10$ blue cells creates a set $S'$ of more than $9$ blue cells in the next step and $R'$ of size $l_1 \times l_2$ is the smallest rectangle which covers $S'$. If $l_1\geq 5$ (or similarly $l_2\geq 5$), then there exist a cell $v$ in $S'$ and the leftmost column of $R'$ and a cell $u$ in $S'$ and the rightmost column of $R'$. Since $Sh(u,v)=0$, we need at least $5$ blue cells in $S \cap \hat{N}(v)$ and $5$ blue cells in $S \cap \hat{N}(u)$. There is a column(s) between these two disjoint blue sets. Therefore, the blue cells which are made by these $10$ blue cells are created by only blue cells in one of these two disjoint blue sets or are created in the \textit{boundary} of two vertices $u,v$ which is defined as $B(u,v):=\{w: w\in \hat{N}(\hat{N}(v))\cap \hat{N}(\hat{N}(u))\}$. We know a blue set of size $5$ can create a blue set of size at most $2$ and $|B(u,v)|\leq 5$. Therefore, these $10$ blue cells create at most $9$ blue cells which means for $s'\geq 10$, we need at least another cell in $S$ which means $s\geq 11$. If $l_1 \leq 4$ and $l_2 \leq 4$, then $l_1=l_2=4$ or $l_1=3$ and $l_2=4$ (similarly $l_1=4$ and $l_2=3$). If $l_1=3$ and $l_2=4$ (similarly $l_1=4$ and $l_2=3$), one can easily check, even a completely blue square of size $3\times 4$ creates only $8$ blue cells one step later. The case of $l_1=l_2=4$ could be checked by a simple computer program\footnote{Notice that all cases cannot be checked by programming because there are roughly ${23^2 \choose 10}\approx 2^{68}$ possibilities.}.
    
Finally, we prove $11$ blue cells cannot create more than $10$ blue cells in the next two steps. Assume a set $S$ of $11$ blue cells create a set $S'$ of more than $10$ blue cells and rectangle $R'$ of size $l_1 \times l_2$ is the smallest rectangle which covers $S'$. If $l_1\geq 6$ (or similarly $l_2\geq 6$), then there exist a cell $v$ in $S'$ and the leftmost column of $R'$ and a cell $u$ in $S'$ and the rightmost column of $R'$. Since $Sh(u,v)=0$, we need at least $5$ blue cells in $S \cap \hat{N}(v)$ and $5$ blue cells in $S \cap \hat{N}(u)$. There are at least two columns between these two disjoint blue sets, which are called the boundary columns. Now, there are two possibilities for the remaining blue cell $w$ ($s=11$). If it is not in the boundary columns, then we will have two independent blue sets of size $6$ and $5$ which can create at most $4$ and $2$ blue cells in the next generation, respectively. In the case that $w$ is in the boundary columns, no blue cell will exist in the boundary columns in the next generation because each vertex in the boundary columns has at most three blue neighbors in $g_0$ without considering $w$. Furthermore, $w$ is in the neighborhood of at most one of the mentioned disjoint blue sets because as we mentioned there are at least two columns in the boundary. If $w$ is in the neighborhood of one of the two disjoint blue sets, it is in the neighborhood of at most three cells in that set because the first and second blue sets are subsets of $\hat{N}(v)$ and $\hat{N}(u)$ (square-shape), respectively. Therefore, each of the disjoint blue sets can make at most $2$ blue cells in the next generation and $w$ can contribute to at most three other blue cells which provides the upper bound of $7$ on the number of blue cells which could be created in this case. The cases of $l_1=l_2=4$ and $l_1=5$ (or similarly $l_2=5$) could be checked by a simple computer program. If $l_1=3$ and $l_2=4$, one can easily check, even a completely blue square of size $3\times 4$ creates only $8$ blue cells one step later. 

Now, we discuss the case of $p_b\gg n^{-\frac{1}{6}}$.  Similar to the proof of Theorem \ref{theorem 30}, consider $\forall 1\leq i'< \lfloor \frac{n}{4} \rfloor-1$ $\wedge$ $\forall 1\leq j'< \lfloor \frac{n}{4} \rfloor-1$ $S_{i',j'}=\{(i,j)| 4(i'-1)+1\leq i\leq 4i' \wedge 4(j'-1)+1\leq j\leq 4j'\}\setminus \{(i,j)|i\in\{4i', 4i'-3\} \wedge j\in\{4j', 4j'-3\}\}$ (see Figure \ref{fig 9} (a)) as $\Theta(n^2)$ disjoint robust sets. Based on Theorem \ref{theorem 15}, majority model in this state reaches a cycle of bichromatic generations with high probability. Actually, by utilizing Theorem \ref{theorem 40} and Corollary \ref{corollary 7}, we can say it reaches a cycle of bichromatic generations of size one or two in $\mathcal{O}(n^2)$ steps.$\Box$

\begin{figure}[h]
\begin{center}
\includegraphics[width=1.0\textwidth]{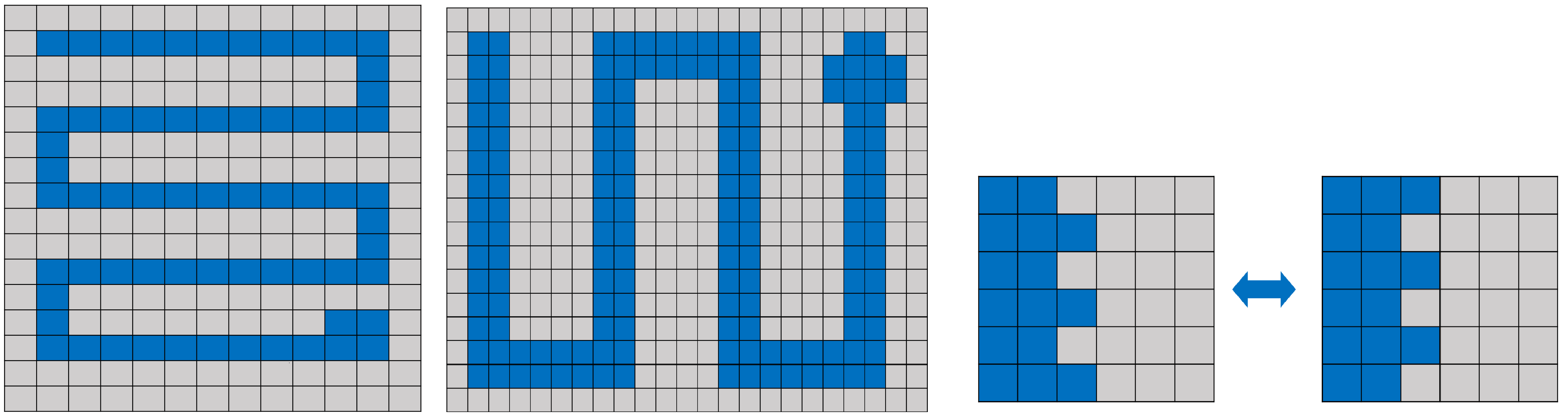}
\caption{(left) An initial generation which needs $\Theta(n^2)$ steps to stabilize in Neumann neighborhood (middle) in Moore neighborhood (right) a two-periodic configuration \label{fig 12}}
\end{center}
\end{figure}

The presented $\mathcal{O}(n^2)$ bound on the consensus time of the process in Theorem \ref{theorem 30} and Theorem \ref{theorem 29} are tight up to a constant factor because there are some initial generations which need $\Theta(n^2)$ steps to stabilize (see Figure \ref{fig 12} (a) and (b): inspired by an example in \cite{balister2010random}). Furthermore, in addition to one-periodic configurations (for instance, a monochromatic generation), two-periodic configurations also can occur (see Figure \ref{fig 12} (c)).

\subsection{Biased Majority Cellular Automata}
\label{biased majority}
In this section, we prove that biased majority cellular automata show two phase transitions in Neumann neighborhood; more accurately, we prove there are two thresholds $0<p_1,p_2<1$ so that $p_b\ll p_1$, $p_1\ll p_b \ll p_2$, and $p_2 \ll p_b$ result in r-monochromatic generation, stable coexistence of both colors, and b-monochromatic generation, respectively. The threshold values depend on the size of the smallest b-eternal and r-eternal sets because they play a critical role in the final status of the process. The proof of the first phase transition is built on the existence or non-existence of a blue eternal set in the initial generation, but for the second phase transition, we need a more complicated argument and we also exploit prior results by Schonmann \cite{schonmann1990finite}.

The most surprising point is that alternating the tie-breaking rule changes the model's behavior substantially. In the case of majority cellular automata, the initial concentration of blue color must be very close to $1$ to guarantee the final complete occupancy by blue with high probability, but in the case of biased majority cellular automata, even initial concentration very close to zero can result in b-monochromatic generation almost surely. More precisely, in majority case with Neumann neighborhood, only $p_r \ll n^{-1/2}$ (which implies $p_b$ is almost one for large $n$) results in final complete occupancy by blue with high probability while in the biased case, $1/\sqrt{\log n} \ll p_b$ outputs a fully blue generation almost surely. It seems that the intuition behind this drastic change is the significant change in the size of the smallest r-eternal set. Roughly speaking, by changing the tie-breaking rule from conservative to biased, the size of the smallest r-eternal set switches from a small constant to linear size\footnote{In the torus $T_{n,n}=(V,E)$ with Neumann neighborhood and the biased majority model, an r-eternal set $S$ must have a vertex in every two consecutive columns or every two consecutive rows; otherwise, in a generation $g$ where $g|_S=r$ and $g|_{V\setminus S}=b$, there are two consecutive fully blue rows and two consecutive fully blue columns which result in a b-monochromatic generation after at most $\mathcal{O}(n^2)$ steps which implies the size of the smallest r-eternal set is at least $\lfloor n/2\rfloor$.} in terms of $n$ in the torus $T_{n,n}$ while the size of the smallest b-eternal set reduces to a smaller constant.    

To prove biased majority cellular automaton with Neumann neighborhood shows a threshold behavior with two phase transitions, we need some results form previous sections (like Corollaries \ref{theorem 40}, \ref{corollary 7}, and Theorem \ref{theorem 15}) and also the following theorem (Theorem \ref{theorem 3}) which was proved by Schonmann \cite{schonmann1990finite}.

\begin{theorem}\cite{schonmann1990finite}
\label{theorem 3}
In the biased majority model and the torus $T_{n,n}$ with Neumann neighborhood, $1/\sqrt{\log n} \ll p_b$ results in final complete occupancy by blue color with high probability.
\end{theorem}

\begin{theorem}
\label{theorem 2}
In the biased majority model, the torus $T_{n,n}$ with Neumann neighborhood has two phase transitions i.e. with high probability:
\\
(i) $p_b \ll n^{-1}$ results in final complete occupancy by red in constant number of steps 
\\
(ii) $n^{-1} \ll p_b \ll 1/\sqrt{\log n}$ guarantees to reach a cycle of bichromatic generations of length one or two in $\mathcal{O}(n^2)$ steps
\\
(iii) $1/\sqrt{\log n} \ll p_b$ outputs  b-monochromatic generation in $\mathcal{O}(n^2)$ number of steps.
\end{theorem}
\textbf{Proof:} We prove parts (i), (ii), and (iii) one by one as follows.

(i) Let random variable $X$ denote the number of blue cells in generation $g_1$. By considering $p_b \ll n^{-1}$ and the fact that a vertex needs at least two blue cells in its neighborhood in $g_0$ to be blue in $g_1$, we have:
\[
E[X]\leq n^2 {4 \choose 2} p_{b}^{2}=o(1).
\] 
By utilizing Markov's Inequality, one can easily see that with high probability $X=0$ which implies $g_1$ is red monochromatic.

(ii) We will show that both colors red and blue almost surely will survive forever; then based on Corollaries \ref{theorem 40} and \ref{corollary 7}, the process reaches a cycle of bichromatic generations of period one or two in $\mathcal{O}(n^2)$ steps.

\begin{figure}[h]
\begin{center}
\includegraphics[width=0.2\textwidth]{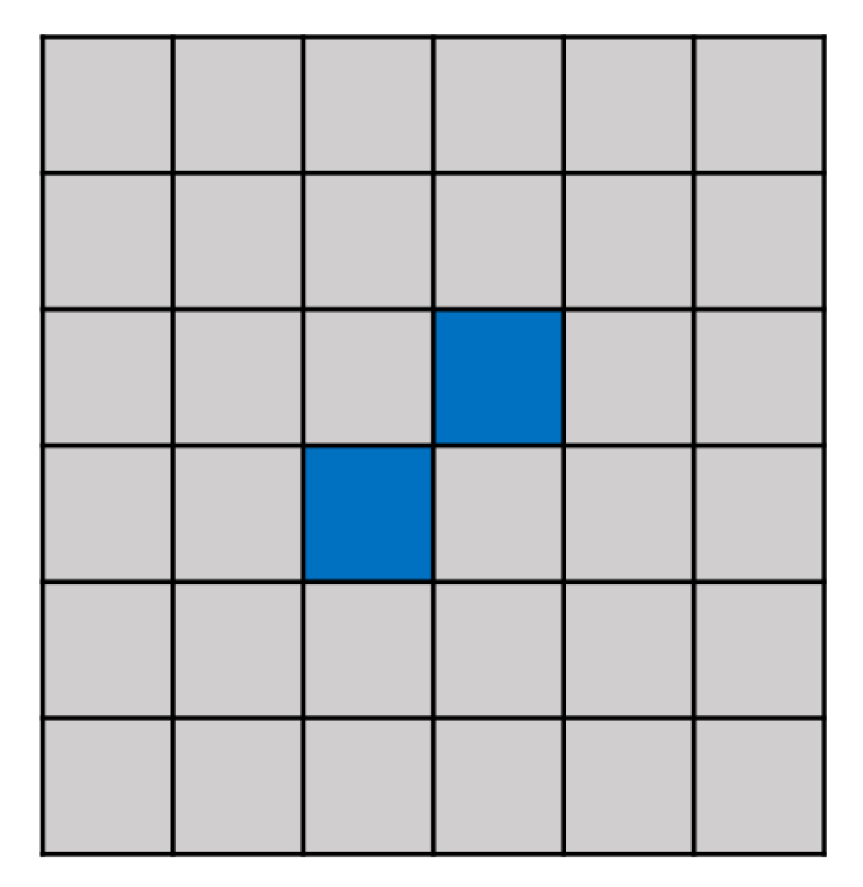}
\caption{A b-eternal set in the biased majority model and $T_{n,n}$ with Neumann neighborhood \label{fig 1}}
\end{center}
\end{figure}

First, we show blue color survives in all upcoming generations with high probability. One can easily see that set $S=\{\{(2i,2j),(2i+1,2j+1)\}: 0\leq i,j< \lfloor n/2\rfloor-1\}$ (see Figure \ref{fig 1}) contains $\Theta(n^2)$ disjoint b-eternal sets of size two in $T_{n,n}$. Now easily by utilizing Theorem \ref{theorem 15} and $n^{-1} \ll p_b$, one can show that with high probability there exists at least a blue eternal set in $g_0$ which guarantees the survival of blue color.

Before we prove that for $p_b \ll 1/\sqrt{\log n}$, red color will never die with high probability, we need to prove the following proposition.

\textbf{Proposition 2}:\textit{ In the biased majority model, the torus $T_{n,n}=(V,E)$ with Neumann neighborhood, and $p_b \ll 1/\sqrt{\log n}$, there is no connected blue component of size $\log n$ in $g_1$ with high probability.}

First, we prove that the number of connected components of size $\log n$ which include a fixed vertex $v$ is at most $16^{\log n}$; since there are $n^2$ number of vertices in the torus $T_{n,n}$, the number of connected components of size $\log n$ is bounded by $n^2 16^{\log n}$. Every connected component of size $\log n$ which includes vertex $v$ has a spanning tree of size $\log n$, rooted at $v$; in turn, each tree rooted at $v$ identifies a unique connected component in the torus $T_{n,n}$. Thus, the number of trees of size $\log n$ rooted at $v$ is an upper bound for the number of connected components of size $\log n$ which include $v$. Now, each such tree can be encoded with $\log n$ 4-bit numbers, where each 4-bit number specifies for the nodes (in DFS order, say) which children (top, right, down, left) are present.


Furthermore, the probability that a connected component $S$ of size $\log n$ is blue in generation $g_1$ is at most ${5\log n \choose \log n/5}p_{b}^{\log n /5}$ because $S$ needs at least $\log n/5$ blue vertices in $\hat{N}(S)$ in $g_0$. That is true because each blue vertex in $g_0$ can create at most $5$ blue vertices in $g_1$ (very generously) and only the color of vertices in $\hat{N}(S)$ in generation $g_0$ impact the color of vertices in set $S$. Notice that $|\hat{N}(S)|\leq 5 \log n$.

Let random variable $Z$ denote the number of connected blue components of size $\log n$ in $g_1$. Now, by using Stirling's approximation (for $1 \leq k \leq n$, ${n \choose k}\leq (ne/k)^k$) \cite{romik2000stirling} and $p_b \ll 1/\sqrt{\log n}$, we have:
\[
E[Z]= \mathcal{O}(n^216^{\log n}) {5\log n \choose \log n/5}p_{b}^{\log n /5} = \mathcal{O}(n^216^{\log n}) (25e)^{\log n/5} o(1/\sqrt{\log n})^{\log n/5}=o(1)
\]      
which implies with high probability there is no connected blue component of size $\log n$ in $g_1$ by utilizing Markov's Inequality \cite{feller1968introduction}, and it finishes the proof of Proposition 2.
 
Now, we prove that for $p_b \ll 1/\sqrt{\log n}$ red color will never die, with high probability. Consider the blue connected components in generation $g_1$, and assume set $M$ is the set which contains the smallest rectangles covering these blue connected components (for more details, see the proof of Theorem \ref{theorem 29} and as we mentioned, the smallest covering rectangle for a set is not necessarily unique, but in this proof considering any smallest covering rectangle for a set works). Now, we run Algorithm \ref{algorithm} (Rectangulation Procedure) by considering $M$ as the input. We remind that for two rectangles $R_1,R_2 \subseteq V$, $d(R_1,R_2):=\min_{u\in R_1,v\in R_2}d(u,v)$, which is called the distance of $R$ and $R'$, where for two vertices (cells) $u,v \in V$, $d(u,v)$ is the size of the shortest path between $u$ and $v$ in terms of the number of edges minus one (for instance the distance between two adjacent vertices is zero and we define $d(v,v)=0$ for a vertex $v$) in the torus $T_{n,n}=(V,E)$ with Neumann neighborhood. The algorithm always terminates because in the worst case, it outputs $M'=\{R\}$ such that $R$ is a rectangle which covers the whole torus. We claim the algorithm terminates before reaching this situation with high probability which means $M'$ (the output) includes at least two disjoint rectangles. If our claim is true, then red color will survive forever because blue cells are clustered in some squares which are surrounded by red vertices. Definitely, blue cells in a rectangle cannot exceed the rectangle in all upcoming generations because the distance between each pair of rectangles is at least two i.e. each red cell out of the rectangles has at least two red neighbors; therefore, all red vertices out of these rectangles will survive forever.

Now, we prove our claim i.e. the process terminates before reaching a rectangle which covers the whole torus. In the above procedure, in each step two rectangles whose distance is less than two are combined. One can see that if the process combines two rectangles $R$ and $R'$ respectively of size $l_1\times l_2$ and $l'_{1}\times l'_2$, the new rectangle's size is not larger than $(2\max (l_1,l_1')+1)\times (2\max (l_2,l'_2)+1)$. Hence, if the process wants to reach the case in which $M'$ contains only one rectangle covering the whole torus, $M'$ at some step must contain a rectangle $R_c$ of size $l_3\times l_4$ so  that $\log n \leq l_3 \leq 3\log n$ and $l_4 \leq 3 \log n$ (or similarly $\log n \leq l_4 \leq 3\log n$ and $l_3 \leq 3 \log n$). This is true because the size of a new rectangle provided by the combining process in terms of width and length can be at most three times larger than the largest rectangle in the previous step and based on Proposition 2, we know that the size of all initial rectangles (in input $M$) is smaller than $\log n\times\log n$ with high probability; then we should pass by such an aforementioned rectangle to reach a rectangle which covers the whole torus. Furthermore, $R_c$ contains at least $\lfloor\max (l_3,l_4)\rfloor/4$ blue cells which are mutually in distance at least two from each other because based on the combining process, $R_c$ contains at least a blue cell in every two consecutive columns (rows) intersecting $R_c$. Without loss of generality assume $\max (l_3,l_4)=l_3$ and let call the columns intersecting $R_c$ from left to right consecutively $c_{1}^{\prime},\cdots,c_{l_3}^{\prime}$, then if we consider a blue cell in the intersection of $R_c$ and  every other pair of columns (i.e. a blue cell in $(c_{4i-1}^{\prime}\cup c_{4i}^{\prime})\cap R_c$ for $1 \leq i \leq \lfloor l_3/4\rfloor$), we take at least $\lfloor l_3/4 \rfloor$ blue cells which are mutually in distance at least two from each other. Now, we prove there is no such a rectangle in $g_1$ with high probability which demonstrates that the algorithm terminates before reaching a rectangle of length (or width) larger than $3 \log n$ which consequentially implies that the aforementioned claim is correct. To prove that, let random variable $X$ denote the number of rectangles in $T_{n,n}$ which have the aforementioned properties for $R_c$ in generation $g_1$. Firstly, every cell is contained in at most $\Theta(\log^4n)$ number of rectangles of the desired size ($\log n \leq l_3 \leq 3\log n$ and $l_4 \leq 3 \log n$) which implies there are $\Theta(n^2\log^4n)$ such rectangles. Furthermore, the probability that a rectangle of size $c_1\log n \times c_2\log n$ for $c_1\geq c_2$ contains $\lfloor c_1\log n/4\rfloor$ blue cells in $g_1$ which are mutually in distance at least two from each other is (generously) bounded by ${(c_1\log n)^2\choose \lfloor c_1 \log n/4\rfloor}({4 \choose 2}p_{b}^{2})^{\lfloor c_1\log n/4\rfloor}$ because each of these blue cells independently needs at least two blue cells in $g_0$ in its neighborhood to become blue in $g_1$. Finally, by considering $c$ as a constant, we have:
\[
E[X] = \mathcal{O}(n^2\log^4n){(c\log n)^2 \choose \lfloor(c\log n)/4\rfloor}({4 \choose 2}p_{b}^{2})^{\lfloor(c\log n)/4\rfloor}
\]
Now, again by using Stirling's approximation and $p_b \ll 1/\sqrt{\log n}$ we have:
\[
E[X] = \mathcal{O}(n^2\log^4n)({4\choose 2}5ec\log n)^{\lfloor(c\log n)/4\rfloor}o((1/\log n)^{\lfloor(c\log n)/4\rfloor})= o(1).
\] 
which finishes the proof of part (ii).
  
(iii) By Theorem \ref{theorem 3}, we know that $1/\sqrt{\log n} \ll p_b$ results in final complete occupancy by blue with high probability. Furthermore based on Corollary \ref{corollary 7}, the consensus time of the process is $\mathcal{O}(n^2)$. $\Box$

Similar to the examples in Figure \ref{fig 12} in the case of majority model, we can show that the presented bounds on the periodicity and consensus time of biased majority cellular automata also are tight up to a constant.

\subsection*{Conclusion}
In the present paper, we analyzed and proved some properties regarding the behavior of two very fundamental majority-based rules on a graph $G=(V,E)$, especially a torus which corresponds to a cellular automaton with (biased) majority rule. First, we presented some results regarding the consensus time and periodicity of both majority and biased majority models. Then, we introduced two basic concepts of robustness and eternalness. Building on our results about these two concepts, periodicity, and consensus time, and exploiting some other techniques like rectangulation, we showed majority and biased majority cellular automata show a threshold behavior with two phase transitions.  

As we discussed, the value of the thresholds in both models depend on the size of the smallest robust set and the smallest eternal set. For instance, in a torus and majority model (majority cellular automaton) $|V|^{-1/r_s}$ is the threshold where $r_s$ is the size of the smallest robust set. It is a natural question whether a threshold (whose value depends on $r_s$) can be obtained for a larger class of graphs, for instance a sub-class of expander graphs or vertex-transitive graphs. Similarly, this question might also be investigated for the case of the biased majority model. 

However, the first step to answer the aforementioned question might be to study the behavior of both models for d-dimensional tori ($d>2$). Roughly speaking, we believe most of the techniques presented in this paper could be applied to higher dimensions, but probably the rectangulation technique needs to be adapted in the way that could be utilized in whatever dimension.

Another interesting subject which might be taken into consideration is the relationship between the concept of connectivity and threshold values. Intuitively speaking, it sounds there is a direct relation between connectivity and the thresholds; for instance, in the majority model on a complete graph, the threshold is almost $1/2$, and a small positive deviation from $1/2$ for a color is sufficient to win with high probability, but in an isolated graph (all vertices are isolated i.e. there is no edge) the threshold is almost zero, and a color with a constant initial probability, even very close to one, has a small chance of final complete occupancy. For instance, we will observe that in majority cellular automata by switching from Neumann neighborhood to Moore neighborhood (doubling the number of edges) the thresholds all get bigger. 

At the end, we would like to introduce two other interesting variants of majority-based models which could be points of interest in future research.

Given an initial generation $g_0$ such that $\forall v\in V$, $Pr[g_0(v)=b]=p_b$ and $Pr[g_0(v)=r]=1-p_b$ independently of all other vertices. In \emph{Random Majority Model} $\forall i\geq 1$ and $v\in V$, $g_i(v)$ is equal to the color that occurs most frequently in $N(v)$ in $g_{i-1}$, and in case of a tie, $v$ chooses among red and blue uniformly at random. More formally: 
\[
g_i(v) =\left\{\begin{array}{lll}\uar\{r,b\}, &\mbox{if $|N^{g_{i-1}}_b(v)|=|N^{g_{i-1}}_r(v)|$}, \\
\argmax_{c\in\{b,r\}} |N^{g_{i-1}}_c(v)|,&\mbox{otherwise}
\end{array}\right..
\]
\emph{Conservative Majority Model} is exactly our original majority model except we change the neighborhood model from $N$ to $\hat{N}$ which means $\forall i\geq 1$ and $v\in V$, $g_i(v)$ is equal to the color that occurs most frequently in $\hat{N}(v)$ in $g_{i-1}$, and in case of a tie, $v$ conserves its current color. More Formally:
\[
g_i(v)= \left\{\begin{array}{lll}g_{i-1}(v), &\mbox{if $|\hat{N}^{g_{i-1}}_b(v)|=|\hat{N}^{g_{i-1}}_r(v)|$}, \\
\argmax_{c\in\{b,r\}} |\hat{N}^{g_{i-1}}_c(v)|,&\mbox{otherwise}
\end{array}\right..
\]

\bibliographystyle{acm} %
\bibliography{ColorWar}
\newpage     
\end{document}